\documentclass[aps,prl,reprint,superscriptaddress]{revtex4-1}
\usepackage{graphicx}
\usepackage[rightcaption]{sidecap}
\usepackage{braket}
\usepackage[utf8]{inputenc}
\usepackage{hyperref}
\usepackage{amsmath} 
\usepackage{amssymb}  
\usepackage{amstext}  
\usepackage{graphicx}
\usepackage{amssymb}
\usepackage{color}
\usepackage{enumerate}

\begin{document}

\pagestyle{plain}

\title{Intrinsic limit to electron spin coherence in InGaAs quantum dots featuring strain-induced nuclear dispersion}

\author{R. Stockill}
\email[Electronic address: ]{rhjs2@cam.ac.uk}
\affiliation{Cavendish Laboratory, University of Cambridge, JJ Thomson Avenue, Cambridge CB3 0HE, United Kingdom}

\author{C. Le Gall}
\affiliation{Cavendish Laboratory, University of Cambridge, JJ Thomson Avenue, Cambridge CB3 0HE, United Kingdom}

\author{C. Matthiesen}
\affiliation{Cavendish Laboratory, University of Cambridge, JJ Thomson Avenue, Cambridge CB3 0HE, United Kingdom}

\author{L. Huthmacher}
\affiliation{Cavendish Laboratory, University of Cambridge, JJ Thomson Avenue, Cambridge CB3 0HE, United Kingdom}

\author{E. Clarke}
\affiliation{EPSRC National Centre for III-V Technologies, University of Sheffield, Sheffield, S1 3JD, UK}

\author{Maxime Hugues}
\affiliation{CNRS-CRHEA, rue Bernard Grégory, 06560 Valbonne, France}

\author{M. Atat{\"u}re}
\email[Electronic address: ]{ma424@cam.ac.uk}
\affiliation{Cavendish Laboratory, University of Cambridge, JJ Thomson Avenue, Cambridge CB3 0HE, United Kingdom}

\begin{abstract}
The Zeeman-split spin-states of a single electron confined in a self-assembled quantum dot provide an optically-accessible spin qubit.  For III-V materials the nuclear spins of the solid-state host provide an intrinsic noise source, resulting in electron-spin dephasing times of few nanoseconds. While a comprehensive study of electron-spin dynamics at low magnetic field has recently been carried out, what limits the electron coherence in these systems remains unclear, in part due to the dominant effect of measurement-induced dynamic polarisation of the nuclear bath. We develop an all-optical method to access the quantum dot spin-state without perturbing the nuclear environment. We use this method to implement Hahn-echo decoupling and reach the intrinsic limit to coherence set by inhomogeneous strain fields coupling to quadrupolar moments of the nuclear bath. These results indicate that the extension of electron spin coherence beyond this few-microsecond limit necessitates the reduction of strain-induced quadrupolar broadening in these materials.
\end{abstract}
\date{\today}
\maketitle

Indium-gallium-arsenide (InGaAs) quantum dots possess strong, coherent optical transitions [1], enabling ultrafast all-optical control of the spin ground states [2] and their entanglement with single photon states [3-5], key building blocks for networked quantum information. Implementing quantum information schemes beyond elementary protocols will place stringent demands on the retention of spin coherences. Decoupling techniques, such as Hahn-echo and higher-order multi-pulse protocols, extend coherence times by filtering the effects of correlated noise [6,7]. In this way the coherence of solid-state spin-qubit realisations such as the nitrogen-vacancy centre in diamond and phosphorus donors in silicon has been found to be limited by the spin diffusion of a small number of environmental nuclei [8,9]. In contrast, electrons confined to self-assembled InGaAs quantum dots experience a particularly rich nuclear environment: the electron couples via the contact hyperfine interaction to a dense bath of $\sim10^{4}-10^{5}$ nuclear spins. Hahn-echo decoupling has been used to recover spin coherence in self-assembled quantum dots for times up to 2.6 $\mathrm{\mu s}$ [10,11,12], however a full explanation of the loss of coherence has been lacking. Understanding the intrinsic processes limiting coherence is essential to exploiting the highly desirable optical properties of these structures.

Self-assembled quantum dots typically form by strain-driven nucleation during epitaxial growth, resulting in inhomogeneous electric field gradients within the crystal lattice. Another source of local field gradients is the random alloying of atomic species in these systems. The nuclear spins in the quantum dot couple to these gradients via quadrupolar moments, resulting in spatially-dependent shifts of their energy levels [13,14]. In this way strain isolates individual nuclear spins such that the bath remains coherent longer than for unstrained systems [15] and maintains polarisation for multiple hours [16]. At the same time, the interaction between the electron spin and the strained bath presents unique challenges, namely the non-linear measurement-induced back-action of nuclei, often observed as the dragging of resonance frequencies by detuning resonant optical probes [17]. This process affects both the evolution of electron spin states and the detuning of resonant spin-readout probes to the extent that it can limit state retrieval and mask the mechanisms influencing the electron-spin dynamics [18].

\begin{figure}
\includegraphics[width=1\columnwidth,angle=0]{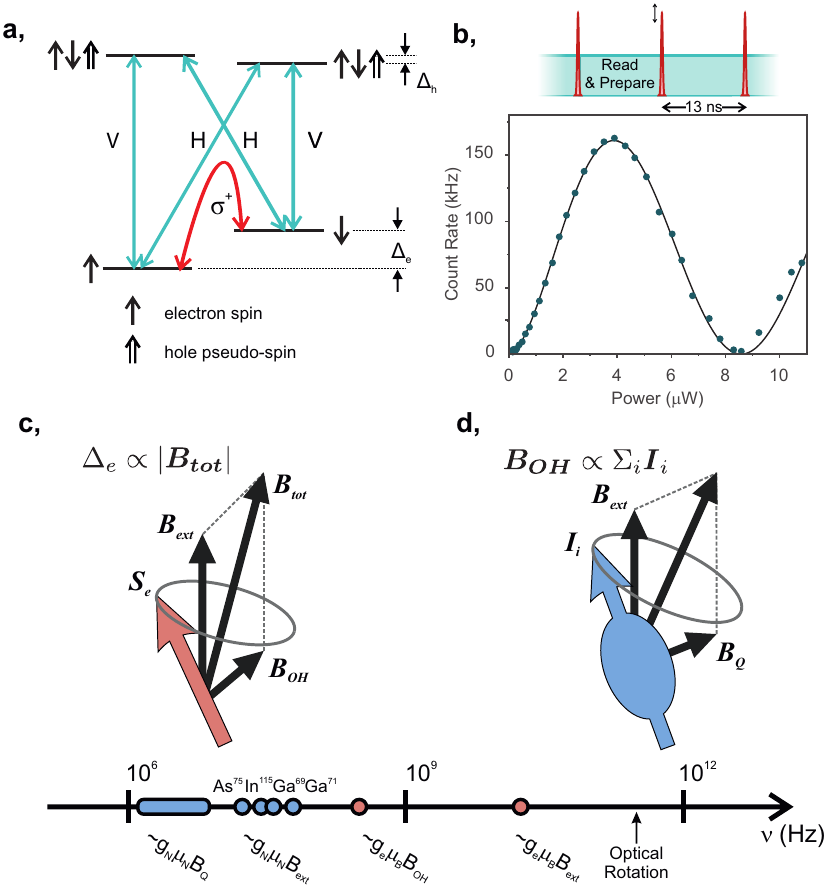}
\caption{ \label{fig1} \textbf{Optical access to InGaAs quantum dot spins and relevant spin dynamics. a,} Energy level structure for a negatively charged quantum dot in Voigt geometry, showing the optically allowed transitions and their respective polarisations. The red arrow represents the effective coherent coupling between the spin ground states, which we create using a $\mathrm{\sigma^+}$- polarised picosecond pulse detuned by $\sim$3 nm from the optical resonance. \textbf{b,} Pulse sequence and recorded count rates for the control of a single electron spin. The depicted scheme contains both the picosecond rotation pulse and the continuous spin readout. The curve in the readout count rate is a sinusoidal fit, with a sublinear power dependence. \textbf{c,} The electron spin splitting is due to both the external magnetic field and the hyperfine coupling with the nuclear bath, represented by the Overhauser field $\mathrm{\mathbf{B_{OH}}}$. \textbf{d,} The nuclear spins are subject to the external field and quadrupolar coupling to electric field gradients, represented here by the effective field $\mathrm{\mathbf{B_{Q}}}$. The frequency axis below displays the relevant frequencies for nuclear and electron dynamics at the few-Tesla external field regime, as well as the optically induced electron spin Rabi-frequency.}
\end{figure}
Optical access to the spin state of an electron resident in an InGaAs quantum dot is provided by the level structure in Fig. 1a. An in-plane magnetic field permits four equal-strength near-infrared transitions to the excited trion states, at 969 nm in our case, and splits the electron-spin ground states by $\mathrm{\Delta_e}$, whilst the excited states are split by $\mathrm{\Delta_h}$. Addressing an optical transition resonantly provides spin-dependent optical readout and prepares a well-defined ground state by optical pumping [19]. Application of a circularly polarised and spectrally detuned pulse rotates the electron spin via the AC-Stark shift [2]. This can be observed in the fluorescence rate as power-dependent spin Rabi oscillations, plotted in Fig. 1b, when picosecond-long rotation pulses are accompanied by a resonant drive for spin readout and repump. Through this we can rotate the electron spin with a fidelity o f$\sim97\%$ within 2 ps.

An electron spin superposition state stored in the quantum dot evolves at the spin splitting $\Delta_e$. This frequency, principally determined by the external quantisation field $\mathrm{\mathbf{B_{ext}}}$, is influenced by the contact hyperfine interaction with the nuclear spins within the electronic wavefunction. This interaction can be represented semiclassically by an effective magnetic field, the Overhauser field $\mathrm{\mathbf{B_{OH}}}$, as depicted in Fig. 1c. In the limit of   $\mathrm{\mathbf{B_{ext}}}$ $\gg$  $\mathrm{\mathbf{B_{OH}}}$, Overhauser field components parallel to the electron quantisation couple linearly to the spin splitting [20], while perpendicular terms perturb the splitting quadratically [21,22]. The high frequency dynamics of the Overhauser field are determined by the nuclear bath precessing in the external field, and also the coupling to electric field gradients via quadrupolar moments, represented in Fig. 1d as an effective field $\mathrm{\mathbf{B_{Q}}}$ [23]. The sum effect of the two is such that the resulting Overhauser field fluctuations contain components both parallel and perpendicular to the external field direction. The frequency chart in Fig. 1 outlines the relevant timescales for each of these processes in a few-Tesla external field. At these fields, the fastest process is electron spin precession at $\mathrm{g_{e}\mu_{e} \sim 6 GHz T^{-1}}$, broadened by the Overhauser field width $\mathrm{g_{e}\mu_{e}B_{OH} \sim}$ 100 MHz [24]. Atomic-species dependent nuclear Zeeman splitting of $\mathrm{g_N\mu_N \sim 10 MHz\ T^{-1}}$ is supplemented by the quadrupolar coupling $\mathrm{g_N\mu_NB_Q \sim}$ 1-10 MHz [25]. The effect of other dynamics, such as dipolar interaction between nuclear spins [15], electron-mediated nuclear flip flops [16] or the precession of nuclei due to hyperfine coupling with the electron [26,27] are neglected in this regime due to their weaker coupling strengths (see supplementary). For reference, we include the Rabi frequency of our spin control at $\sim500$ GHz, up to two orders of magnitude faster than any other process.
\begin{figure*}
\includegraphics[width=1\textwidth,angle=0]{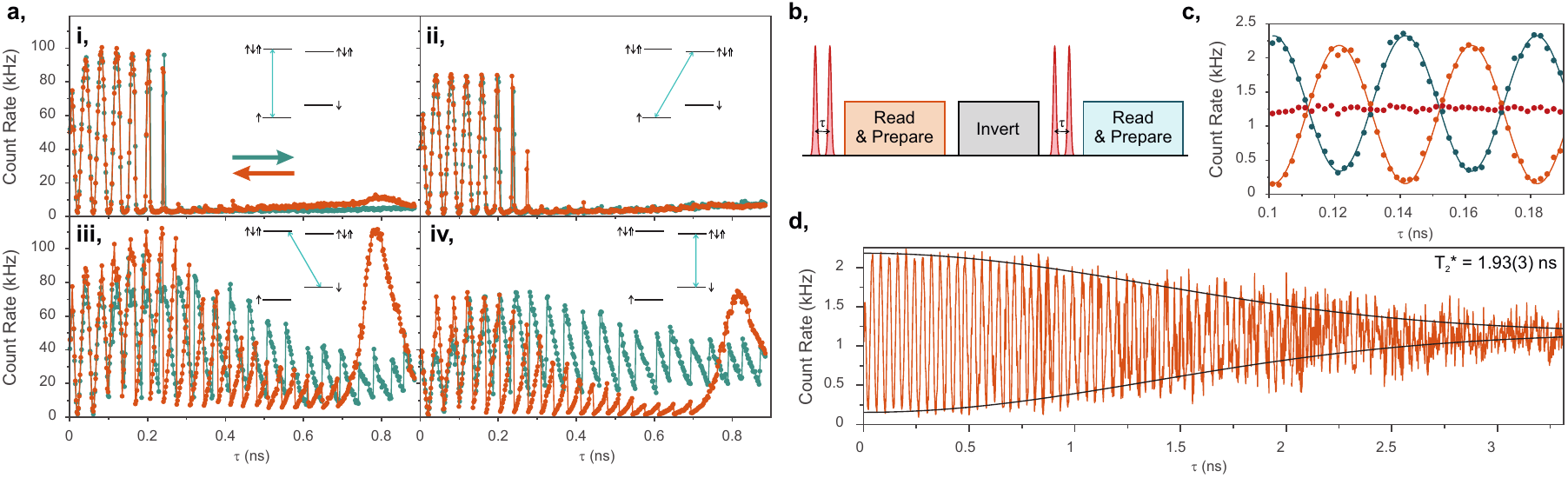}
\caption{ \label{fig2}. \textbf{Nuclear polarisation in Ramsey Interferometry. a,} Readout fluorescence as a function of the delay between $\pi$/2-rotations using the four allowed transitions to probe and prepare the electron spin. The insets indicate the probed transition. The non-sinusoidal shape is due to the polarisation of nuclear spins in the quantum dot. The data are taken for $\mathrm{B_{ext}}$ = 4 T, resulting in an electron spin splitting of 25.2 GHz. \textbf{b}, Alternating pulse sequence to suppress nuclear spin polarisation. Every second rotation sequence begins with an inverted spin state. \textbf{c}, Count rates from the alternating sequence. The two out-of-phase signals with (blue circles) and without (orange circles) a spin inversion produce a time-averaged signal without phase dependence (red circles). The curves are sinusoidal fits to data. \textbf{d}, Full free induction decay with alternating sequence. The probed transition is the same as in panel a, i. The black curve is a Gaussian envelope with a 1.93-ns decay.}
\end{figure*}

The nuclear dynamics introduced in Fig. 1d affect the storage and retrieval of arbitrary spin states from the quantum dot. To investigate the extent of their effect we measure the Ramsey interference of spin-states for varying delay, $\tau$ between two $\pi$/2 spin-rotations. The sub-panels of Fig. 2a, labelled (i-iv), display the resonant spin readout signal when we probe each of the four available transitions (see inset). All data sets show a striking departure from the expected sinusoidal beating associated with the precession of the electron spin. When probing the spin-down projection (iii \& iv), a hysteretic saw-tooth behaviour emerges [18], whilst for probing spin-up projection (i \& ii) the readout signal is suppressed for delays beyond 200 ps. This behaviour emerges from dynamic polarisation of the nuclear spins due to resonant driving of the optical transitions, producing a feedback loop between bath polarisation and the measured electron state [5]. As can be seen in Fig. 2a, it is the ground-state electron projection of the readout transition which dictates the feedback dynamics we observe, consistent with mechanisms based on the non-collinear hyperfine interaction [28]. The effect of this state-dependent polarisation is to frustrate time-averaged measurements of the electron-nuclear system and to prevent access to the full timescale of spin coherence. In order to inhibit the electron-state dependent polarisation of the nuclear bath, we repeat the same measurement as in Fig. 2a, however we invert the spin before every other repeat, as depicted in Fig. 2b. The two resulting measurements presented in Fig. 2c suppress the phase dependence in the average signal (solid red circles). In this way we prevent the build-up of nuclear polarisation and resolve the evolution of the unperturbed electron-nuclear system, as shown in Fig. 2d for one half of the measurement pair. The symmetric, Gaussian-amplitude decay with $T_{2}^*$ = 1.93(3) ns is testament to the large amplitude, yet quasi-static environmental noise, consistent with an Overhauser field standard deviation of 33 mT [3,27] (see supplementary).

Having successfully decoupled nuclear bath polarisation from our resonant optical spin readout, we implement a Hahn-echo scheme to assess the extent to which we can protect electron-spin coherence. The measurement sequence is displayed in Fig. 3a for a 91.2-ns delay between the $\pi$/2-rotations, in which the central $\pi$-rotation refocuses the dephased spin state. We sweep the central rotation by $\tau < T_{2}^*$ and record the visibility of the oscillatory readout signal [10]. In the limit of perfect rotations the visibility is a direct measure of the coherence of the spin state. Figure 3b displays the integrated readout from the pulse sequence in Fig. 3a, which emerges only after having decoupled the readout from bath polarisation as in Fig. 2 (see supplementary). The two traces in Fig. 3b correspond to the echo signal with (blue circles) and without (orange circles) initial inversions.

Figure 3c shows the visibilities we recover when varying storage time, T, up to 1.3 $\mu$s for four different external magnetic fields. The refocussing pulse filters the quasi-static noise components that resulted in the 2-ns time-averaged dephasing, yet the electron state is still susceptible to high-frequency dynamics of the Overhauser field. The first points of interest in Fig. 3c are the collapse of coherence beyond 13 ns at low field (2 T) and the damped oscillatory behaviour at short times for higher fields. These effects are principally due to the linearly coupled Overhauser field components along the electron-spin quantisation direction. The linearly coupled terms arise from the tilt between the effective quadrupolar field axis and the external field. The large strain in these samples results in a quadrupolar coupling strength equivalent to $\sim 1$-T magnetic field. We only observe spin coherence beyond $\sim 13$ ns at relatively high external fields ($>$ 2 T) as the Zeeman interaction starts to dominate nuclear dynamics, reducing linearly coupled Overhauser field components. In the presence of high-frequency noise, Hahn echo becomes a sensitive spectroscopic tool, evidenced by the oscillatory visibility at short storage times and high external fields. This allows us to determine the spectral composition of the nuclear bath at these frequencies by viewing the echo sequence as a filter of the Overhauser field noise spectrum.

\begin{figure*}
\includegraphics[width=1\textwidth,angle=0]{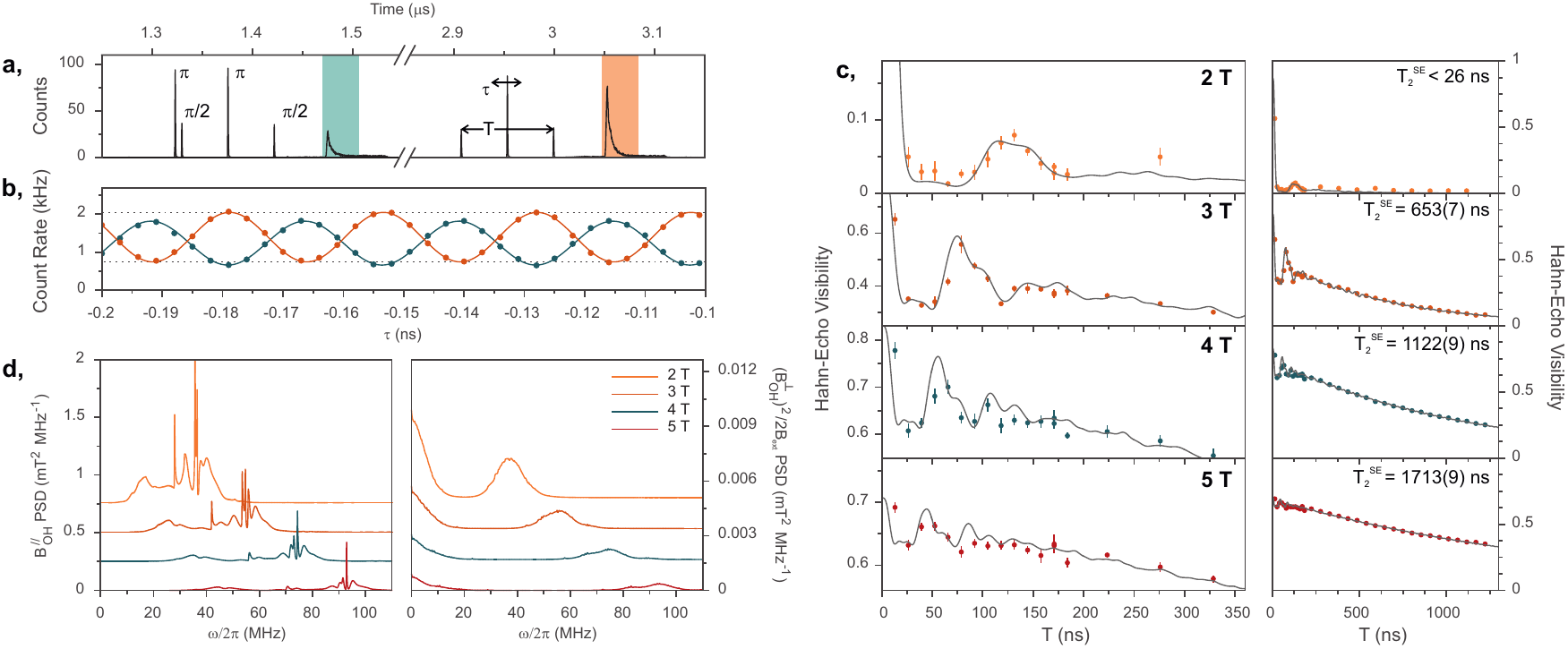}
\caption{ \label{fig2} \textbf{Hahn-echo measurement with suppressed nuclear feedback. a,} Histogram of pulse sequence to perform and measure all-optical Hahn echo for a T = 91.2 ns storage time. The marked areas are regions of interest for spin readout. \textbf{b,} Integrated count rates in regions of interest for the two readout pulses as the central rotation pulse is scanned by $\tau$, at an external field of 3 T, revealing a visibility of 47.6(1.2)\%. The blue trace is of smaller visibility due to imperfect spin inversion. \textbf{c,} Extracted visibilities for varying storage time and external field values. The visibilities are drawn from the readout without initial inversion. Error bars show the standard error of the mean for repeated measurements. The curve is the result of modelling the echo sequence as a spectral filter of the nuclear spectra in (d). Right-hand panels show zoomed-out traces of long storage times, showing the exponential visibility decay and the fitted decay times. \textbf{d,} Power spectral densities of the nuclear noise. The left panel displays the linearly coupled Overhauser field components, along the quantisation direction of the electron. The right panel displays the spectra of quadratically coupled transverse components.}
\end{figure*}
The power spectral densities for both the linearly coupled and transverse quadratically coupled noise components are plotted in the left and right panels of Fig. 3d. These are calculated from the quadrupolar and Zeeman Hamiltonians for the four relevant atomic species, Ga$^{69}$, Ga$^{71}$, As$^{75}$ and In$^{115}$. The broad structures result from integrating over a Gaussian distribution of quadrupolar energies and orientations, corresponding to the inhomogeneous spread of electric field gradients (see supplementary). Our choice of parameters is based on an indium concentration of 0.5 with strain profiles motivated by atomistic calculations [29]. The curves in Fig. 3c are the calculated electron Hahn-echo visibilities resulting from such noise spectra. Fitting the power spectrum amplitude to our echo modulation is consistent with an Overhauser field standard deviation of 28-40 mT. By decomposing the spectra into the constituent atomic species (see supplementary) we find that indium plays the dominant role in the dephasing of the electron, owing to its large nuclear spin, whilst the gallium isotopes have a negligible effect.
For electrostatically confined III-V quantum dots, Hahn-echo coherence has been observed to follow an exp$[-(T/T_2)^4]$ behaviour [22], which is predicted to emerge from the dipolar coupling between nuclear spins [30]. In contrast, the coherence we observe decays exponentially at high fields ($\mathrm{B_{ext}} \geq$ 3 T) and long times (T $>$ 200 ns), at a field-dependent rate extending to 1.7 $\mu$s at 5 T. This decay stems from the quadratically coupled perpendicular Overhauser field components, plotted in the right panel of Fig. 3d. These feature a broad low-frequency shoulder due to the inhomogeneous quadrupolar spread, responsible for the long-time loss of coherence. The perpendicular components couple to the electron splitting with a $\mathrm{1/B_{ext}}$ dependence, which reduces the effective amplitude at higher fields. Indeed, further data taken for a different quantum dot at a higher external field (7 T) shows coherence times extending to 2.7 $\mu$s (see supplementary). This mechanism provides a faster source of dephasing than theoretical predictions for coherence limited by nuclear-spin diffusion in these materials [30], which should extend beyond 6 $\mu$s. Instead, the quadrupolar inhomogeneity is a source of uncorrelated noise, resulting in the characteristic exponential decay of electronic coherence.

By inhibiting dynamic nuclear spin polarisation during resonant optical access to spn states, we observe the quadrupolar interaction of the nuclear bath with inhomogeneous electric field gradients providing an intrinsic bound to spin coherence in InGaAs quantum dots. In other systems, such as electrostatically confined GaAs quantum dots, multi-pulse decoupling schemes can be used to further prolong the spin coherence beyond Hahn-echo limits. Such schemes rely on noise being correlated at least on the order of the pulse separation times [7]. The broadband nature of the quadrupole-coupled nuclear dynamics, however, decorrelates the electron splitting in $\sim$20 ns. Therefore, only pulses spaced on or closer than this timescale could extend spin coherence in these systems. In this way the strain that is necessary for the formation of the quantum dot provides the intrinsic source of dephasing. We note, using a single heavy-hole as an alternative quantum-dot spin qubit should allow for longer dephasing times due to a much weaker hyperfine interaction, although other mechanisms can limit its coherence [31]. Longer coherence times for electron spin qubits could be realised either for systems without quadrupolar nuclear moments, as in II-VI quantum dots, or through strain-free growth methods. One particular example of the latter is provided by optically active GaAs/AlGaAs quantum dots grown by droplet epitaxy. With over two orders of magnitude weaker local strain than the quantum dots used here, significantly different nuclear dynamics have been reported [32]. With reduced quadrupolar broadening absolute coherence times that are competitive with other spin qubit realisations could be achieved, whilst still maintaining the ultrafast control rates and optical integration the material properties enable.

We gratefully acknowledge financial support by the University of Cambridge and the European Research Council ERC Consolidator Grant Agreement No. 617985. C.M. acknowledges Clare College, Cambridge, for financial support through a Junior Research Fellowship. We thank H. Bluhm, T. Botzem, L. Cywinski, D. Kara, M. Stanley and J.M. Taylor for fruitful discussions and S. Topliss for technical assistance.

\section{Methods}

\subsection{Sample}
We use a sample grown by molecular beam epitaxy containing a single layer of self-assembled InGaAs quantum dots in a GaAs matrix, embedded in a Schottky diode for charge state control. The Schottky diode structure comprises an n$^{+}$-doped layer 35 nm below the quantum dots and a $\sim$6-nm thick partially transparent titanium layer evaporated on top of the sample surface. This device structure allows for deterministic charging of the quantum dots and shifting of the exciton energy levels via the DC Stark effect. 20 pairs of GaAs/AlGaAs layers form a distributed Bragg reflector below the quantum dot layer for increased collection efficiency in the spectral region between 960 nm and 980 nm. Spatial resolution and collection efficiency are enhanced by a zirconia solid immersion lens in Weierstrass geometry positioned on the top surface of the device. The device is cooled in a liquid-helium bath cryostat to 4.2 K and surrounded by a superconducting magnet.

\subsection{Spin inversion to prevent nuclear spin polarisation}
The inversion necessary to cancel phase terms in the average readout signal and suppress nuclear polarisation can be provided by either a coherent $\pi$-rotation, or an incoherent re-pumping step. For Hahn echo a $\pi$-rotation suffices, however for measurements of the time-averaged dephasing, the enhanced sensitivity for longer delays between the $\pi$/2-rotations requires complete spin inversion such that the rotation has to be supplemented with a pumping step.

\subsection{Pulse sequence and detection}
Optical pulse sequences are constructed from a Ti:Sapphire pulsed laser in picosecond-mode, detuned from the optical resonance by $\sim$ 3 nm and a resonant continuous-wave diode laser. Both are modulated with fibre-coupled waveguide electro-optic modulators. The modulators are locked to the 76-MHz repetition rate of the pulsed laser via a pulse delay generator with 8-ps jitter. Additional suppression of the readout pulse is provided by an acousto-optical modulator, realising 6000:1 suppression of readout lasers and $>$2000:1 rotation pulse suppression. The readout laser is used at a power below optical saturation to avoid spin-pumping when not reading the spin state. The readout fluorescence is filtered from the resonant laser by polarisation mode rejection. Additional filtering of the detuned rotation pulses is provided by a holographic grating with a 30-GHz full width at half maximum and a first-order diffraction efficiency above 90\%. The photon detection events are recorded with a time-correlation unit and a single photon detector with a timing resolution of 350 ps.

\section{References}
\begin{enumerate}[{[1]}]

\item	Matthiesen, C., Vamivakas, A. N. \& Atatüre, M. Subnatural Linewidth Single Photons from a Quantum Dot. Phys. Rev. Lett. \textbf{108}, 93602 (2012).

\item Press, D., Ladd, T. D., Zhang, B. \& Yamamoto, Y. Complete quantum control of a single quantum dot spin using ultrafast optical pulses. Nature \textbf{456}, 218–221 (2008).

\item	Gao, W. B., Fallahi, P., Togan, E., Miguel-Sanchez, J. \& Imamoglu, A. Observation of entanglement between a quantum dot spin and a single photon. Nature \textbf{491}, 426–430 (2012).

\item	De Greve, K. et al. Quantum-dot spin-photon entanglement via frequency downconversion to telecom wavelength. Nature \textbf{491}, 421-425 (2012).

\item	Schaibley, J. R. et al. Demonstration of quantum entanglement between a single electron spin confined to an InAs quantum dot and a photon. Phys. Rev. Lett. \textbf{110}, 1–5 (2013).

\item	Hahn, E. L. Spin Echoes. Phys. Rev. \textbf{80}, 580–594 (1950).

\item	Viola, L., Knill, E. \& Lloyd, S. Dynamical Decoupling of Open Quantum Systems. Phys. Rev. Lett. \textbf{82}, 2417–2421 (1999).

\item	Abe, E. et al. Electron spin coherence of phosphorus donors in silicon: Effect of environmental nuclei. Phys. Rev. B \textbf{82}, 9–12 (2010).

\item	de Lange, G., Wang, Z. H., Ristè, D., Dobrovitski, V. V \& Hanson, R. Universal dynamical decoupling of a single solid-state spin from a spin bath. Science \textbf{330}, 60–63 (2010).

\item	Press, D. et al. Ultrafast optical spin echo in a single quantum dot. Nature Photon. \textbf{4}, 367–370 (2010).

\item	Greilich, A. et al. Mode locking of electron spin coherences in singly charged quantum dots. Science \textbf{313}, 341–345 (2006).

\item	Xu, X. et al. Coherent Population Trapping of an Electron Spin in a Single Negatively Charged Quantum Dot. Nature Phys. \textbf{4}, 2–5 (2008).

\item	Bechtold, A. et al. Three-stage decoherence dynamics of an electron spin qubit in an optically active quantum dot. Nature Phys. \textbf{11}, 1005-1008 (2015).

\item	Hackmann, J., Glasenapp, P., Greilich, A., Bayer, M. \& Anders, F. B. Influence of the Nuclear Electric Quadrupolar Interaction on the Coherence Time of Hole and Electron Spins Confined in Semiconductor Quantum Dots. Phys. Rev. Lett. \textbf{115}, 207401 (2015).

\item	Chekhovich, E. a., Hopkinson, M., Skolnick, M. S. \& Tartakovskii,  a. I. Quadrupolar induced suppression of nuclear spin bath fluctuations in self-assembled quantum dots. Nature Commun. \textbf{6}, 6348 (2014).

\item	Latta, C., Srivastava, A. \& Imamoglu, A. Hyperfine interaction-dominated dynamics of nuclear spins in self-assembled InGaAs quantum dots. Phys. Rev. Lett. \textbf{107}, 1–5 (2011).

\item	Latta, C. et al. Confluence of resonant laser excitation and bidirectional quantum-dot nuclear-spin polarization. Nature Phys. \textbf{5}, 758–763 (2009).

\item	Ladd, T. D. et al. Pulsed nuclear pumping and spin diffusion in a single charged quantum dot. Phys. Rev. Lett. \textbf{105}, 1–4 (2010).

\item	Xu, X. et al. Fast spin state initialization in a singly charged InAs-GaAs quantum dot by optical cooling. Phys. Rev. Lett. \textbf{99}, 1–4 (2007).

\item	Botzem, T., McNeil, R. P. G., Schuh, D., Bougeard, D. \& Bluhm, H. Quadrupolar and anisotropy effects on dephasing in two-electron spin qubits in GaAs. at http://arxiv.org/abs/1508.05136 (2015)

\item	Cywiński, Ł., Witzel, W. M. \& Das Sarma, S. Pure quantum dephasing of a solid-state electron spin qubit in a large nuclear spin bath coupled by long-range hyperfine-mediated interactions. Phys. Rev. B \textbf{79}, 245314 (2009).

\item	Bluhm, H. et al. Long coherence of electron spins coupled to a nuclear spin bath. Nature Phys. \textbf{7}, 10 (2010).

\item	Sinitsyn, N. a., Li, Y., Crooker, S. a., Saxena,  a. \& Smith, D. L. Role of nuclear quadrupole coupling on decoherence and relaxation of central spins in quantum dots. Phys. Rev. Lett. \textbf{109}, 1–5 (2012).

\item	Braun, P.-F. et al. Direct Observation of the Electron Spin Relaxation Induced by Nuclei in Quantum Dots. Phys. Rev. Lett. \textbf{94}, 116601 (2005).

\item	Chekhovich, E. A. et al. Structural analysis of strained quantum dots using nuclear magnetic resonance. Nature Nanotech. \textbf{7}, 646–650 (2012).

\item	Stanley, M. J. et al. Dynamics of a mesoscopic nuclear spin ensemble interacting with an optically driven electron spin. Phys. Rev. B \textbf{90}, 195305 (2014).

\item	Merkulov, I. a \& Rosen, M. Electron spin relaxation by nuclei in semiconductor quantum dots. Phys. Rev. B \textbf{65}, 1–8 (2002).

\item	Högele,  a. et al. Dynamic nuclear spin polarization in the resonant laser excitation of an InGaAs quantum dot. Phys. Rev. Lett. \textbf{108}, 1–5 (2012).

\item	Bulutay, C. Quadrupolar spectra of nuclear spins in strained In xGa 1-xAs quantum dots. Phys. Rev. B \textbf{85}, 1–12 (2012).

\item	Witzel, W. M. \& Das Sarma, S. Quantum theory for electron spin decoherence induced by nuclear spin dynamics in semiconductor quantum computer architectures: Spectral diffusion of localized electron spins in the nuclear solid-state environment. Phys. Rev. B \textbf{74}, 035322 (2006).

\item	De Greve, K. et al. Coherent control and suppressed nuclear feedback of a single quantum dot hole qubit. Nature Phys. \textbf{7}, 5 (2011).

\item	Sallen, G. et al. Nuclear magnetization in gallium arsenide quantum dots at zero magnetic field. Nature Commun. \textbf{5}, 3268 (2014).

\end{enumerate}

\end{document}